\documentclass[a4paper,12pt]{article}
\usepackage{graphicx}

\begin{document}

\title{Pair creation by a photon
in a strong magnetic field}
\author{V. N. Baier
and V. M. Katkov\\
Budker Institute of Nuclear Physics,\\ Novosibirsk, 630090, Russia}

\maketitle

\begin{abstract}
The process of pair creation by a photon in a strong magnetic field
is investigated basing on the polarization operator in the field.
The total probability of the process is found in a relatively simple
form. The probability exhibits a "saw-tooth" pattern because of
divergences arising when the electron and positron are created at
threshold of the Landau energy levels. The pattern  will be washed
out at averaging over any smooth photon energy distribution. The new
results are obtained in the scope of the quasiclassical approach: 1)
in the case when the magnetic field  $B \ll B_0,~ (B_0$ is the
critical field) the new formulation extends the photon energy
interval to the case when the created particles are not
ultrarelativistic; 2) the correction to the standard quasiclassical
approximation  is found showing the range of applicability of the
approach at high photon energy as well. The very important
conclusion is that for both cases $B \ll B_0$ and $B \geq B_0$ the
results of the quasiclassical calculation are very close to averaged
probabilities of exact theory in a very wide range of photon
energies. The quasiclassical approximation is valid also for the
energy distribution if the electron and positron are created on
enough high levels.

\end{abstract}

\newpage

\section{Introduction}

Study of QED processes in a strong external field is stimulated
essentially by existence of very strong magnetic fields in nature.
There is a series of evidences for the existence of neutron stars
possessing magnetic fields close the critical field strength
$B_0=m^2/e=(m^2c^3/e\hbar)=4.41\cdot 10^{13}$~G. The rotating
magnetic dipole model, in which the pulsar loses rotational energy
through the magnetic dipole radiation, was confirmed with the
discovery that the spin-down power predicted for many pulsars is in
a quite good energetic agreement with the observed radiation, and
gives magnetic fields $B \sim 10^{11}-10^{13}$~G \cite{Ru}. There
are presently around 1600 spin-powered radio pulsars.

Another class of neutron stars was discovered at X-ray and
$\gamma$-ray energies and may possess even stronger surface magnetic
fields $B \sim 10^{14}-10^{15}$~G. Such stars are now referred to as
magnetars \cite{DuT}, since they most probably derive their power
from their magnetic fields rather than from spin-down energy loss.
Two different classes of objects are thought to be magnetars: the
Soft Gamma-ray Repeaters (SGR), discovered as sources of short
bursts of hard X-rays ($>30$ keV) with super-Eddington luminosity
and the Anomalous X-ray Pulsars (APS), discovered as persistent,
soft ($<10$ keV) X- ray sources with pulsations of several seconds
and spinning-down on time scales of $10^{4}-10^{5}$ years. An
increasing number of common properties, pointing to a close
relationship of these two apparently different classes of objects
and leading to their interpretation as magnetars, has been found
(see e.g.\cite{Me}).

Pair creation by a photon in a strong magnetic field is the basic
QED reaction which can play the significant role in processes in
vicinity of a neutron star.

Investigation of pair creation by a photon in a strong magnetic
field was started in 1952 independently by Klepikov and Toll
\cite{Kle1, To}. In Klepikov's paper \cite{Kle2}, which was based on
the solution of the Dirac equation in a constant and homogeneous
magnetic field, the probabilities of radiation from an electron and
$e^-e^+$ pair creation by a photon were obtained for magnetic field
of arbitrary strength on the mass shell\footnote{We use the system
of units with $\hbar=c=1$ and the metric
$ab=a^{\mu}b_{\mu}=a^0b^0-{\bf a}{\bf b}$} ($k^2=0,~k$ is the
4-momentum of photon). In 1971 Adler \cite{Ad} calculated the photon
polarization operator in the mentioned magnetic field using the
proper-time technique developed by Schwinger \cite{Schwi} and
Batalin and Shabad \cite{Bash} calculated the photon polarization
operator in a constant and homogeneous electromagnetic field for
$k^2\neq 0$ using the Green function in this field found by
Schwinger \cite{Schwi}. In 1975 Strakhovenko and present authors
calculated the contribution of a charged-particles loop with $n$
external photon lines having applied the proper-time method in a
constant and homogeneous electromagnetic field \cite{BKS}. The
explicit expression for the photon polarization operator was found.
In a purely magnetic field ({\bf E}=0) the imaginary part of the
polarization operator, which determines the probability of the pair
creation in general case ($k^2\neq 0$), was analyzed in detail  near
the root singularities originating at each threshold of $e^-e^+$
pair creation in the state with the given Landau quantum numbers.
This peculiarity of the pair creation process was observed by
Klepikov \cite{Kle2}. Shabad \cite{Sha} performed an extensive
analysis of the photon polarization operator found in \cite{Bash} in
a purely magnetic field. It was shown that its imaginary part in the
limit $k^2 \rightarrow 0$ reduces to the result of Klepikov's
calculation \cite{Kle2}. Daugherty and Harding \cite{Daha} analyzed
in detail the total probabilities (the attenuation coefficients) of
the pair creation by a polarized photon in the superstrong field $B
\sim B_0$ basing on the calculations of Toll and Klepikov  and
paying the special attention to the threshold region $\omega \geq
2m$, where the probabilities exhibit a "saw-tooth" pattern due to
the mentioned root singularities.

For high Landau levels Klepikov \cite{Kle2} (and Toll) found the
quasiclassical representations for the probabilities of radiation
and pair creation in the case $B \ll B_0$. In 1968 the present
authors developed the operator quasiclassical method for the
consideration of electromagnetic processes in an external field
\cite{BK} which is valid also in nonuniform and (or) depending on
time field under conditions: 1) $B \ll B_0$, 2) the charged
particles have relativistic energies. Both the radiation and pair
creation by a photon were considered. The method is given in detail
in \cite{BKF, BKS1}. Tsai and Erber \cite{Tse} deduced the
quasiclassical probability of pair creation using the photon
polarization operator in a magnetic field calculated by Adler
\cite{Ad}.

In the Sec.2 the exact probability of the pair creation by a photon
was obtained for general case $k^2 \neq 0$ starting from the
polarization operator in a magnetic field. The technical details of
the derivation are given in Appendix A. In Sec.3 the standard
quasiclassical approximation (SQA) is outlined in the limit
$\mu\equiv B/B_0 \ll 1,~r\equiv\omega^2/4m^2 \gg 1$, where $\omega$
is the photon energy. The quasiclassical characteristics of the
process depend on the parameter $\kappa=2\mu\sqrt{r}$. The
correction to SQA is found. Details of calculation of the correction
starting from the polarization operator in a magnetic field is given
in Appendix C. If the photon energy is not high, so that $\mu \ll
r-1 \ll \mu^{-2}$ and SQA is not applicable, the new quasiclassical
approach (at low energy near threshold (th)) is developed in
Appendix B. One can expect that the photon energy distribution in a
vicinity of neutron star is wide and smooth. At averaging over this
distribution the "saw-tooth" pattern will be washed out. It is very
important that the result the averaging is very close to SQA even at
$\mu \geq 1$. The corresponding analysis has been carried out in
Sec.3. The energy spectrum of created pair is discussed in Sec.4.
Here SQA is very useful also. A remark concerning derivation of the
energy spectrum in SQA is given in Appendix D.

\section{Probability of pair creation by a photon: exact theory}

Our analysis is based on the expression for the polarization
operator obtained in \cite{BKS}, see Eqs.(3.19), (3.33). For pure
magnetic field ({\bf E}=0) this polarization operation can be
written in the diagonal form
\begin{equation}
\Pi^{\mu\nu}=-\sum_{i} \kappa_i\beta_i^{\mu}\beta_i^{\nu},\quad
\beta_i\beta_j=-\delta_{ij}, \quad \beta_ik=0, \quad
\sum_i\beta_i^{\mu}\beta_i^{\nu}=\frac{k^{\mu}k^{\nu}}{k^2}-g^{\mu\nu},
 \label{d0}
\end{equation}
where
\begin{eqnarray}
&& \beta_1^{\mu}=\frac{k^2 k_{\perp}^{\mu}+{\bf k}_{\perp}^2
k^{\mu}}{k_{\perp}\sqrt{k^2(\omega^2-k_3^2)}}, \quad
\beta_2^{\mu}=\frac{F^{\mu\nu}k_{\nu}}{Bk_{\perp}}, \quad
\beta_3^{\mu}=\frac{F^{\ast\mu\nu}k_{\nu}}{B\sqrt{\omega^2-k_3^2}},
\nonumber \\
&& \kappa_1=\Omega_1(r-q),\quad \kappa_2=\kappa_1-q\Omega_2, \quad
\kappa_3=\kappa_1+r\Omega_3,
\nonumber \\
&& r=\frac{\omega^2-k_3^2}{4m^2}, \quad q=\frac{k_{\perp}^2}{4m^2},
\quad r-q=\frac{k^2}{4m^2}, \label{d1}
\end{eqnarray}
where the axis 3 directed along the magnetic field {\bf B}, ${\bf
k}_{\perp}{\bf B}=0$, $k_{\perp}=\sqrt{{\bf k}_{\perp}^2}$, $\omega$
is the photon energy, $F^{\mu\nu}$ is the tensor of electromagnetic
field, $F^{\ast\mu\nu}$ is the dual tensor of electromagnetic field,
\begin{equation}
\Omega_i=-\frac{\alpha
m^2}{\pi}\int\limits_{-1}^1dv\int\limits_0^{\infty-i0}f_i(v,x)\exp(i\psi(v,x))dx.
\label{d2}
\end{equation}
Here
\begin{eqnarray}
&& f_1(v,x)=\frac{\cos vx}{\sin x}-v\frac{\cos x \sin vx}{\sin^2 x},
\quad f_2(v,x)=2\frac{\cos vx-\cos x}{\sin^3x}-f_1(v,x),
\nonumber \\
&& f_3(v,x)=(1-v^2)\cot x -f_1(v,x),\quad \psi(v,x)=z\frac{\cos x
-\cos vx}{\sin x}-b(v)x, \label{d3}
\end{eqnarray}
where
\begin{equation}
z=\frac{2q}{\mu},\quad b(v)=\frac{1-r(1-v^2)}{\mu}, \quad
\mu=\frac{B}{B_0} \label{d4}.
\end{equation}
Let us note that the positive $x$ axis in the integral Eq.(\ref{d2})
is turned slightly down, and in the function $\Omega_1$ in the
integral over $x$ the subtraction at $\mu=0$ is implied.

The real part of the polarization operator defines the dispersive
properties of the space region with magnetic field. At $r < 1~(b >
0)$ one can turn the integration contour over $x$ to the negative
imaginary axis ($x \rightarrow -ix$) so that
\begin{equation}
\Omega_i=\frac{\alpha
m^2}{\pi}\int\limits_{-1}^1dv\int\limits_0^{\infty}if_i(v,-ix)\exp\left(-bx-z\frac{\cosh
x -\cosh vx}{\sinh x}\right)dx. \label{d5}
\end{equation}
The functions $\Omega_i$ in Eq.(\ref{d5}) are real. These functions,
studied in \cite{BMS} at $\omega < 2m$ and arbitrary value of $\mu$,
define the index of refraction $n_{2,3}$. On mass shell $k^2=0$ one
has
\begin{equation}
 n_{2,3}^2=1-\frac{\kappa_{2,3}}{\omega^2}.\label{d6}
\end{equation}
For the mode 2 the photon polarization vector ${\bf e}_2$ is
perpendicular to the plane formed by the vectors ${\bf B}$ and ${\bf
k}$ and for mode 3 the photon polarization vector ${\bf e}_3$ lies
in this plane ($n_2 \rightarrow n_{\parallel}$ and $n_3 \rightarrow
n_{\perp}$ in Adler's notation \cite{Ad}).

At $r > 1$ the polarization operator eigenvalue $\kappa_i$ acquires
an imaginary part, which determines the probability of $e^-e^+$ pair
creation per unit length $W_i$ by a photon with a given polarization
\begin{equation}
W=\frac{1}{\omega}~{\rm Im}e^{\mu}e^{\nu\ast}~\Pi_{\mu\nu}.
\label{d4a}
\end{equation}
Details of calculation basing on Eqs.(\ref{d1}),(\ref{d2}) are given
in Appendix A. Using $e^{\mu}_{i}=\beta^{\mu}_{i}$ we get the
explicit expressions for $W_i~(i=1,2,3)$ at $k^2\neq 0$:
\begin{equation}
 W_i=-\frac{{\rm Im}\kappa_i}{\omega}=\frac{2\alpha m\mu e^{-z}}{\omega\lambda_c \sqrt{r}}
 \sum_{n,m}\left(1-\frac{\delta_{n0}}{2}\right)\vartheta(g)\frac{z^n k!}{\sqrt{g}~l!}
 d_i,
 \label{d7}
\end{equation}
where $\delta_{n0}$ is the Kronecker symbol, $\vartheta(g)$ is the
Heaviside function: $\vartheta(g)=1$ for $g \geq 0$,~
$\vartheta(g)=0$ for $g < 0$,
\begin{eqnarray}
&& l=\frac{m+n}{2},\quad k=\frac{m-n}{2},\quad
g=r-1-m\mu+\frac{n^2\mu^2}{4r},
\nonumber \\
&& l_{max}=[d(r)],\quad d(r)=\frac{2(r-\sqrt{r})}{\mu}. \label{d8}
\end{eqnarray}
Here $[d(r)]$ is the integer part of $d(r)$,
\begin{eqnarray}
&& d_1=\left(\frac{r}{q}-1\right)G,\quad d_2=\frac{r}{q}G+4\mu
l\vartheta(k-1)L_{k-1}^{n+1}(z)L_{k}^{n-1}(z),
\nonumber \\
&&d_3=\left(1+ \frac{m\mu}{2}-\frac{n^2\mu^2}{4r}\right)F+2\mu
l\vartheta(k-1) L_{k}^{n}(z)L_{k-1}^{n}(z),
\nonumber \\
&& G=\left(\frac{m\mu}{2}-\frac{n^2\mu^2}{4r}\right)F-2\mu
l\vartheta(k-1) L_{k}^{n}(z)L_{k-1}^{n}(z),
\nonumber \\
&&F=\left(L_{k}^{n}(z)\right)^2+\vartheta(k-1)\frac{l}{k}
\left(L_{k-1}^{n}(z)\right)^2, \label{d9}
\end{eqnarray}
where $L_{k}^{n}(z)$ is the generalized Laguerre polynomial, $z$ is
defined Eq.(\ref{d4}). Obtained probabilities $W_i$ agree with found
by Shabad \cite{Sha} but presented in much more compact form.

On the mass shell ($k^2=0$) in Eqs.(\ref{d7})-(\ref{d9}) one has to
put $r=q$. In this case $W_1=0$ and only two polarizations $i=2$ and
$i=3$ remain.  The probability of pair creation averaged over the
photon polarizations acquires especially simple form (in the frame
where $k_3=0$)
\begin{eqnarray}
&& W=\frac{W_2+W_3}{2}=\frac{\alpha e^{-z}}{\lambda_c}
 \sum_{n,m}\left(1-\frac{\delta_{n0}}{2}\right)\vartheta(g)\frac{z^{n-1} k!}{\sqrt{g}~l!}
\nonumber \\
&&\times\left[\left(1+ m\mu-\frac{n^2\mu}{z}\right)F+4\mu
l\vartheta(k-1) L_{k-1}^{n+1}(z)L_{k}^{n-1}(z)\right]
 \label{d10}
\end{eqnarray}

The probabilities of pair creation by a photon Eqs.(\ref{d7}),
(\ref{d10}) contain the factor $1/\sqrt{g}$. The function $g=0$, if
\begin{equation}
 r=\left(\frac{\varepsilon(l)+\varepsilon(k)}{2m}\right)^2,\quad
 \varepsilon(l)=m\sqrt{1+2l\mu},
 \label{d11}
\end{equation}
where $\varepsilon(l)$ is the energy of charged particle on the
Landau energy level. So the probability diverges when the electron
and positron are created on the Landau levels with electron and
positron momentum $p_3=0$, see e.g. \cite{Kle2, BKS}. The origin the
singularity is due the properties of the space volume in the lowest
order of perturbation theory (infinitesimally narrow level). The
reason why this singularity is not a pole but a branch point is that
motion along a field is not quantized.

\section{Probability of pair creation by a photon:\\
quasiclassical approximation}
\setcounter{equation}{0}

In the case $\mu \ll 1,~r-1 \gg 1~(l_{max}\gg 1)$ the mentioned
above divergences are situated very often: $\Delta l \sim \Delta
r/\mu \sim 1$. On the photon energy scale the characteristic
distance between the mentioned peaks will be $\Delta\omega \sim
\omega_0$, where $\omega_0$ is the frequency of motion of created
particle in a magnetic field. This frequency defines the distance
between energy levels in the quasiclassical approximation. So the
resonance properties of pair creation probability become apparent if
the width of energy levels and the effective spread of photon energy
are small comparing with $\omega_0$ value. In the opposite case the
spectrum peculiarities will be washed out and the process
probability will be quite adequate described by approximate
expressions ultimately connected with quasiclassical character of
created particle motion at large Landau quantum numbers.

When the magnetic field is weak comparing with critical one ($\mu
\ll 1$) and for high energy photons ($r \gg 1$) the quasiclassical
approximation was created in \cite{Kle2, BK, Tse}, where the process
probabilities  were derived using the different approaches. In
\cite{Kle2} the asymptotic of Eq.(\ref{d10}) were obtained. The
approach in \cite{BK} was developed for $\mu \ll 1$ and the particle
motion was considered to be quasiclassical (and relativistic) from
the very beginning; it is applicable for nonuniform (and
non-stationary) fields. The pair creation probabilities in the form
Eq.(\ref{c6}) (see Appendix C) were found in \cite{Tse}. Below we
will call it the standard quasiclassical approximation (SQA).

In Appendix C the corrections to SQA were calculated. For
probability of pair creation by unpolarized photon the correction is
\begin{equation}
W^{(1)}=\frac{\alpha m^2\mu^2}{60\sqrt{3}\pi\omega\kappa}
\int\limits_{-1}^{1}\left[2(1+v^2-27z^2)K_{1/3}(z)
+3(7-v^2)zK_{2/3}(z)\right]\frac{dv}{1-v^2}, \label{d12}
\end{equation}
where
\begin{equation}
z=\frac{8}{3(1-v^2)\kappa},\quad \kappa=2\sqrt{r}\mu.
  \label{d13}
\end{equation}
At $\kappa \ll 1$ the main contribution is given by the term
$\propto z^2$, it is
\begin{equation}
W^{(1)}=\frac{\sqrt{6} \alpha m^2\mu^2}{5 \omega\kappa^2}
\exp\left(-\frac{8}{3\kappa}\right), \quad
\frac{W^{(1)}}{W}=-\frac{32\mu^2}{15\kappa^3}.\label{d14}
\end{equation}
It is seen that at $\kappa \ll 1$ the SQA is applicable if
$\mu^2/\kappa^3 \ll 1$.

At $r \leq \mu^{-2/3}$ SQA becomes inapplicable. In this region one
can use the process probability found in Appendix B, which is valid
if $\mu \ll r-1 \ll \mu^{-2}$. Using Eq.(\ref{b5}) we get for the
probability of pair creation by unpolarized photon
\begin{equation}
W^{(th)}=\frac{\alpha
m^2\mu}{4\omega}\frac{3r-1}{\sqrt{r(r-1)l(r)\beta(r)}}
\exp\left(-\frac{\beta(r)}{\mu}\right),
 \label{d15}
\end{equation}
where
\begin{equation}
l(r)=\ln \frac{\sqrt{r}+1}{\sqrt{r}-1},\quad
\beta(r)=2\sqrt{r}-(r-1)l(r).
 \label{d16}
\end{equation}
For large $r \gg 1$, limited by the condition $r \ll
\mu^{-2}~(\kappa \ll 1)$, we have
\begin{equation}
W^{(th)}\simeq\frac{3\alpha m^2\mu}{8\omega}\sqrt{\frac{3r}{2}}
\exp\left(-\frac{4}{3\mu\sqrt{r}}-\frac{4}{15\mu
r^{3/2}}\right)=W^{(SQA)}\exp\left(-\frac{4}{15\mu r^{3/2}}\right)
 \label{d17}
\end{equation}
At $\kappa^3 \gg \mu^2$ Eq.(\ref{d17}) agrees with Eq.(\ref{d14}).
The last expression has essentially wider region of applicability
than pure $W^{(SQA)}$.

The behavior of pair creation probability for unpolarized photon at
$\mu=0.1 (1/\mu \gg 1)$ as a function of $r$ is illustrated in
Fig.1. The probability in saw-tooth form is calculated according to
Eq.(\ref{d10}). It is seen that the characteristic distance between
peaks is of the order $\mu=0.1$. The result of averaging over this
interval $(-\mu/2+r \div \mu/2+r)$ is given by the dashed line which
is smoothed out all peculiarities of original behavior and which is
in a very good agreement with the thick curve calculated according
to Eq.(\ref{d15}).

At $\kappa \geq 1$ Eq.(\ref{d15}) becomes inapplicable. However in
this case SQA is valid. From Eq.(\ref{c6}) we have the probability
of pair creation by unpolarized photon
\begin{equation}
W^{(SQA)} =\frac{\alpha m^2}{3\sqrt{3}\pi\omega}
\int\limits_{-1}^{1}\frac{9-v^2}{1-v^2}K_{2/3}(z)dv,
 \label{d18}
\end{equation}
where $z$ is defined in Eq.(\ref{d13}).  The correction to this
probability is given by Eq.(\ref{d12}), its relative value at
$\kappa \geq 1$ is $\leq \mu^2$. The ratio $W^{(th)}$ and
$W^{(SQA)}$ is shown in Fig.2. The curves 2 and 3 are for polarized
photons, the thick curves is for unpolarized photon (see
Eq.(\ref{d15}) and Eq.(\ref{d18})). The point where the ratio
attains 1 is the boundary of the applicability region of the
corresponding approximation.

The relative contribution of the correction Eq.(\ref{d12}) to SQA at
$\mu=3$ is shown in Fig.3. The upper curve is the probability of
pair creation by a photon in SQA Eq.(\ref{d18}), The lower curve is
the sum $W^{(SQA)}+W^{(1)}$ (the correction is negative at low $r$).
It is seen that the correction is small in the region of SQA
applicability $r \gg \mu/2=1.5$.

It is very important that the quasiclassical approximation appears
to be applicable also for super-strong  magnetic field $\mu \geq 1$.
This is true if the characteristic parameter $l_{max} \gg 1$. It is
assumed in this case that parameter $\kappa$ is large
$(\kappa=2\sqrt{r}\mu \gg 1)$. In the limit $\kappa \gg 1$ the
asymptotic expansion of the probability Eq.(\ref{d18}) is
\begin{equation}
W^{(SQA)} =\frac{\alpha m^2}{\pi\omega}
\left(B\kappa^{2/3}-\frac{2\pi}{3}+\ldots\right),\quad B=\frac{
3^{7/6}5}{14}\frac{\Gamma^3(2/3)}{\Gamma(1/3)}=1.1925..
 \label{d19}
\end{equation}
The asymptotic expansion of correction Eq.(\ref{d12}) is
\begin{equation}
W^{(1)} =\frac{\alpha m^2}{\pi\omega}\frac{\mu^2}{60\sqrt{3}\kappa}
\left(A\kappa^{1/3}-6\pi+\ldots\right),\quad A=\frac{
3^{1/3}2}{5}\frac{\Gamma^3(1/3)}{\Gamma(2/3)}=8.191..
 \label{d20}
\end{equation}
The relative value of the correction defines the boundary of region
where SQA is valid for the super-strong  magnetic field $\mu \geq 1$
\begin{equation}
\frac{W^{(1)}}{W^{(SQA)}} = 3^{-7/3}\frac{7}{125}
\frac{\Gamma^4(1/3)}{\Gamma^4(2/3)}\frac{\mu^2}{\kappa^{4/3}}
\left(1-\frac{6\pi}{A\kappa^{1/3}}+\ldots\right) \simeq
0.066\frac{\mu^2}{\kappa^{4/3}}
\left(1-\frac{2.30}{\kappa^{1/3}}\right)
 \label{d21}
\end{equation}
It is seen from this expression that at $\mu \geq 1$ the
applicability of quasiclassical approximation is controlled by the
parameter $(\mu/2r)^{2/3}=l_{max}^{-2/3}\ll 1$. So, both in the case
$\mu \ll 1$ and in the case $\mu \geq 1$ the quasiclassical
approximation is valid when the particles of pair are created in
states with large Landau quantum numbers. It should be noted
concerning Eq.(\ref{d21}) the smallness of numerical factor and
relatively large contribution of the second term $\propto
\kappa^{-1/3}$. This means that the correction is small even at low
$l_{max}$, and one have to use the exact expression Eq.(\ref{d12}).
Under these circumstances one can use Eq.(\ref{d18}) with the
correction Eq.(\ref{d12}) starting from $r \sim \mu$. This is
illustrated in Fig.4 where the probability of pair creation is shown
at $\mu=3$. The probability in saw-tooth form is calculated
according to Eq.(\ref{d10}). It is seen that the characteristic
distance between peaks is of the order $\mu$ (here $\mu=3$) just as
in the case shown in Fig.1 where the probability for $\mu=0.1$ was
presented. The result of averaging over this interval $(-\mu/2+r
\div \mu/2+r)$ is given by the dashed line which is smoothed out all
peculiarities of original behavior and which is in a very good
agreement with the thick curve which is quasiclassical approximation
calculated according to Eqs.(\ref{d18}) and (\ref{d12}).

\section{Energy distribution of created pair}
\setcounter{equation}{0}

The energy distribution of created pair is symmetric with respect to
exchange of electron and positron, i.e. it is symmetric respect
energy $\omega/2$ and one can consider this distribution in interval
$x=\varepsilon_-/\omega \geq 1/2$. In the frame where $k_3=0$ the
energy of electron and positron is
\begin{equation}
\varepsilon_-(l)=\sqrt{p_3^2+m^2(1+2\mu l)}, \quad
\varepsilon_+(k)=\sqrt{p_3^2+m^2(1+2\mu k)}, \quad
\omega=\varepsilon_-(l)+\varepsilon_+(k).
 \label{s1}
\end{equation}
From Eq.(\ref{s1}) follows that $p_3^2/m^2=g,~(g$ is defined in
Eq.(\ref{d8})). In the plane $(k, l)$ the allowed states of electron
and positron are bounded by the axes and the curve $p_3^2=0$ which
is parabola with the nearest point to origin $l_n=k_n=[(r-1)/2\mu]$.
When $k=0$ the maximal value $l=l_{max}~(l_{max}$ is defined in
Eq.(\ref{d8})). Substituting $p_3^2/m^2$ in $\varepsilon_-$ we get
\begin{equation}
x=\frac{\varepsilon_-}{\omega}=\frac{1}{2}
\left(1+\frac{n\mu}{2r}\right),\quad n=l-k.
 \label{s2}
\end{equation}
So the distribution over energy of created particle is of the
discrete character. The probability $W(n)$ is defined by
Eq.(\ref{d10}), where one has to perform summation at fixed $n=l-k$.
The expression for spectrum contains the factor $1/\sqrt{g}$
discussed above (see Eq.(\ref{d11})).

The energy distribution over electron energy in SQA (see, e.g.
\cite{BKS1}, Eq.(3.50)) is
\begin{equation}
\frac{dW^{(SQA)}}{dx}=\frac{\alpha m^2}{\sqrt{3}\pi \omega}
\left[\frac{x^2+(1-x)^2}{x(1-x)}K_{2/3}(\xi)+
\int_{\xi}^{\infty}K_{1/3}(y)dy\right],
 \label{s3}
\end{equation}
where
\begin{equation}
\xi=\frac{2}{3\kappa x(1-x)}.
 \label{s4}
\end{equation}

At low values of the parameters $\mu$ and $\kappa$ the energy
distribution of pair creation probability has a sharp peak when
$\varepsilon_-=\varepsilon_+=\omega/2$ (for details see Sec.3.2 in
\cite{BKS1}). When $\mu \ll 1$ and $\kappa \geq 1$ the SQA is not
valid on an edge of the energy spectrum where the created particle
is no more relativistic ($1-x \sim m/\omega$). However at smaller
energy ($1-x \ll 1/\kappa= m/(\mu\omega$)) the differential
probability is suppressed exponentially, so the contribution of the
interval $1-x \sim m/\omega$, where SQA is not valid, is negligible
small.

For the super-strong field $\mu \geq 1$ in the region of
applicability of the quasiclassical approximation ($r \gg \mu$) the
parameter $\kappa=2\mu\sqrt{r} \gg 1$. In this case the upper limit
of electron energy $x_b \simeq 1 - 1/(2\sqrt{r})$ is lower (or of
the same order) than the electron energy in the maximum of the
energy distribution of Eq.(\ref{s3}) ($x_m \sim 1 - 1.6/\kappa$).
The region of applicability of the quasiclassical approximation near
$x_b$ is defined by the quasiclassical character of positron
(electron) transverse motion ($p_3=0, k \gg 1$)
\begin{equation}
(1-x)^2-\frac{1}{4r} \gg \frac{\mu}{2r}; \quad 1-x \sim \frac{\Delta
n}{n_m} \gg \frac{1}{\sqrt{n_m}},\quad  \Delta n \gg \sqrt{n_m},
 \label{s5}
\end{equation}
where  $n_m = l_{max}$ Eq.(\ref{d8}).

In Fig.5 the spectrum of created particles is shown for
$\mu=3,~r=110~(n_m=66)$. The spectrum in saw-tooth form is
calculated according to Eq.(\ref{d10}). The dashed curve is the
spectrum averaged over the interval $(-2\mu+r \div 2\mu+r)$. The
thick curve calculated using Eq.(\ref{s3}). It is seen that this
curve is in a quite good agreement with the result of averaging.

In this case the argument $\xi \ll 1$ in Eq.(\ref{s3}) and one can
perform the expansion ($k^3=0$)
\begin{equation}
\frac{dW^{(SQA)}}{dx}=\alpha m \mu
\left[C_1\frac{x^2+(1-x)^2}{(\kappa x(1-x))^{1/3}}+
\frac{1}{3\kappa}-C_2\frac{1}{(\kappa x(1-x))^{5/3}}+\ldots\right],
 \label{s6}
\end{equation}
where
\begin{equation}
C_1=\frac{3^{1/6}}{2\pi}\Gamma(2/3)=0.2588\ldots, \quad
C_2=\frac{3^{-1/6}}{4\pi}\Gamma(1/3)=0.1775\ldots
 \label{s7}
\end{equation}
In Fig.6 the energy distribution of the created electron(positron)
at $\mu=3$ is shown for different r: $r=30,~n_m=16$ (the upper
curve), $r=110,~n_m=66$ (the middle curve), $r=500,~n_m=318$ (the
lower curve), calculated according with Eq.(\ref{s3}) (the solid
curve) and according with Eq.(\ref{s6}) (the dashed curve). Very
good agreement of corresponding curves is seen in the region of SQA
applicability ($x \geq n_m^{-1/2},(1-x) \geq n_m^{-1/2}$).

\section{Conclusion}

In this paper we study the process of pair creation by a photon in a
strong magnetic field basing on the polarization operator in the
field calculated by different methods in a set of papers \cite{Ad,
Bash, BKS}. By using the decomposition of functions into series
containing the Bessel function (see Eq.(\ref{a4})) we deduced the
imaginary part of the polarization operator (which is the process
probability) into the relatively simple combination of the
generalized Laguerre polynomial (see Eq.(\ref{d10})). This
probability exhibits a "saw-tooth" pattern because of the factor
$1/\sqrt{g}$. This pattern  will be washed out at averaging over any
smooth photon energy distribution.

The range of applicability of the quasiclassical approximation
extended in the photon energy interval where the created particles
are no more ultrarelativistic (see Appendix B). Found correction to
standard quasiclassical approximation (SQA) permitted to extend the
range of applicability of the quasiclassical approximation to the
region $r \sim \mu$ at $\mu \geq 1$. From the performed analysis
follows the remarkable conclusion: for both $\mu \ll 1$ and $\mu
\geq 1$ the results of the quasiclassical calculation are very close
to averaged probabilities of exact theory in very wide interval of
photon energies. For the total probability of pair creation the
quasiclassical approach is valid if (see Eq.(\ref{d8}))
\begin{equation}
l_{max}=\left[\frac{2(r-\sqrt{r})}{\mu}\right] \gg 1
 \label{co1}
\end{equation}
In relatively weak field ($\mu \ll 1$) it is valid not far from
threshold: $(r-1)/\mu \gg1$. For this energy interval the relative
correction to the probability Eq.(\ref{d15}) is of the order
$l_{max}^{-1/2}$. In superstrong field $\mu \geq 1$, for
relativistic motion of the particles of the created pair and if
$l_{max} \simeq 2r/\mu \gg 1$ the SQA is valid. The expansion
parameter in this case is $l_{max}^{-2/3}$ (see Eq.(\ref{c11})).

For such field and energy it is helpful to compare the probability
of pair creation by a photon $W$ with the radiation length $L_{rad}$
for photon emission process found in \cite{BKS2}. Conserving the
main terms of the decomposition over the Landau energy number $l$
(see Eq.(\ref{d11})) in the SQA as well as in the first correction
we get ($k^3=0$)
\begin{equation}
L_{rad}^{-1}=\frac{a_{\gamma}}{L}\left[\frac{1}{l^{1/6}}+
\frac{b_{\gamma}}{l^{5/6}} \right],\quad
W=\frac{a_{p}}{L}\left[\frac{1}{l_{max}^{1/6}}+\frac{b_{p}}{l_{max}^{5/6}}
\right],
 \label{co2}
\end{equation}
where
\begin{eqnarray}
\hspace{-12mm}&&
\frac{1}{L}=\alpha^{3/2}\left(\frac{B}{e}\right)^{1/2},\quad
\alpha=\frac{e^2}{\hbar c}=\frac{1}{137},
\nonumber \\
\hspace{-12mm}&& a_{\gamma}=\frac{32}{81} (18)^{-1/6}
\Gamma(2/3)=0.33045\ldots,\quad a_{p}=\frac{15}{7}
(18)^{-1/6}\frac{\Gamma^2(2/3)}{\Gamma^2(1/3)}=0.33819\ldots,
\nonumber \\
\hspace{-12mm}&&
b_{\gamma}=\frac{(12)^{-1/3}}{20}\frac{\Gamma(1/3)}{\Gamma(2/3)}=0.043206\ldots,~
b_{p}=\frac{7(12)^{-1/3}}{1125}\frac{\Gamma^4(1/3)}{\Gamma^4(2/3)}=0.041633\ldots,
 \label{co3}
\end{eqnarray}
where the numbers are taken from Eq.(46) in \cite{BKS2} and
Eqs.(\ref{d19})), (\ref{d20})) above. It should be noted that the
physics characteristics in Eq.(\ref{co2})) don't depend on the
electron mass. It's important that in the formula not only
coefficients $a_{\gamma}$ and $a_{p}$ are very close, but the same
properties possess the coefficients $b_{\gamma}$ and $b_{p}$
reflecting proximity of details of mechanisms of radiation and pair
creation by a photon in a magnetic field. One can expect that this
property will be essential for consideration of electron-photon
shower in a magnetic field. Set of kinetic equations for such shower
with the particle polarization taken into account was investigated
in \cite{BKS3, BKS4}.

\vspace{0.5cm}

{\bf Acknowledgments}

The authors are indebted to the Russian Foundation for Basic
Research supported in part this research by Grant 06-02-16226.

\newpage
\setcounter{equation}{0} \Alph{equation}
\appendix

\section{Appendix}
Let us consider the integral in Eq.(\ref{d2})
\begin{equation}
T_i=\int\limits_{-1}^1dv \int\limits_{0}^{\infty-i0}
f_i(v,x)\exp(i\psi(v,x))dx. \label{a1}
\end{equation}
Using the formula
\begin{equation}
\frac{2}{iz}\cot x\frac{d}{dx}e^{i\psi} =
\left(f_2-f_1-\frac{2b}{z}\cot x\right)e^{i\psi},\label{a2}
\end{equation}
and integrating by parts the integral over $x$ in Eq.(\ref{a1}) we
get
\begin{equation}
\int\limits_0^{\infty-i0}f_2e^{i\psi}dx =
\int\limits_0^{\infty-i0}f_1e^{i\psi}dx
+\frac{2}{z}\int\limits_0^{\infty-i0}\left(b\cot x-\frac{i}{\sin^2
x}\right) e^{i\psi}dx \label{a3}
\end{equation}
Using Eq.7.2.4.(27) \cite{BE} we have
\begin{eqnarray}
\hspace{-8mm}&& e^{-it\cos vx}=\sum_{n=0}^{\infty}
\left(2-\delta_{n0}\right)(-i)^n J_n(t)\cos(nvx), \quad
t=\frac{z}{\sin x},
\nonumber \\
\hspace{-8mm}&&  e^{-it\cos vx}~\cos vx=i\frac{d}{dt}e^{-it\cos vx}
= \sum_{n=0}^{\infty} \left(1-\frac{\delta_{n0}}{2}\right)(-i)^{n+1}
\left(J_{n+1}(t)-J_{n-1}(t)\right)\cos(nvx),
\nonumber \\
\hspace{-8mm}&&  e^{-it\cos vx}~\sin
vx=\frac{1}{itv}\frac{d}{dx}e^{-it\cos vx} =
\frac{2i}{t}\sum_{n=1}^{\infty} (-i)^n n J_{n}(t)\sin(nvx),
 \label{a4}
\end{eqnarray}
where $J_n(t)$ is the Bessel function. Taking into account that the
functions $f_i$ and $\psi$ in the integral over $v$ in Eq.(\ref{d2})
are the even functions of $v$, we can perform following
substitutions in sums in Eq.(\ref{a4})
\begin{equation}
\cos(nvx) \rightarrow e^{invx}, \quad  \sin(nvx) \rightarrow -i
e^{invx}.\label{a5}
\end{equation}
So the expression for $T_i$ Eq.(\ref{a1}) can be written as
\begin{equation}
T_i=\sum_{n}T_i^{(n)},\quad  T_i^{(n)}=\int\limits_{-1}^1dv
\int\limits_{0}^{\infty-i0}
F_i^{(n)}(v,x)\exp(ia_n(v)x)dx,\label{a6}
\end{equation}
where
\begin{eqnarray}
\hspace{-8mm}&& F_1^{(n)}=(-i)^n\exp(iz\cot x)\left[\frac{i}{\sin
x}\left(1-\frac{\delta_{n0}}{2}\right)
\left(J_{n+1}(t)-J_{n-1}(t)\right)-\frac{2vn}{z}\cot x
J_{n}(t)\right],
\nonumber \\
\hspace{-8mm}&& F_2^{(n)}=F_1^{(n)}+(-i)^n\exp(iz\cot x)
\frac{2\left(2-\delta_{n0}\right)}{z}\left(b\cot x-\frac{i}{\sin^2x
}\right)J_{n}(t),
\nonumber \\
\hspace{-8mm}&& F_3^{(n)}=(-i)^n\exp(iz\cot x)(1-v^2)\cot x
\left(2-\delta_{n0}\right)J_{n}(t)-F_1^{(n)},
\nonumber \\
\hspace{-8mm}&& a_n(v)=nv-b,
 \label{a7}
\end{eqnarray}
and the function $F_i^{(n)}$ is a periodic function of $x$.

Let us note that at $x\rightarrow -i\infty$ the asymptotic of
$J_n(z/\sin x)$ is
\begin{equation}
J_n\left(\frac{iz}{\sinh x}\right)\simeq J_n(2ize^{-x})\simeq
\frac{(iz)^n}{n!}e^{-nx}.\label{a8}
\end{equation}
In the case $a_n(v) < n$ the integration contour over $x$ in
Eq.(\ref{a6}) can be unrolled to the lower imaginary semiaxis and
the contribution to the imaginary part of $T_i^{(n)}$ vanishes.

One can represent $T_i^{(n)}$ in the form
\begin{eqnarray}
&& T_i^{(n)}=\int\limits_{-1}^1dv \int\limits_{0}^{2\pi}
F_i^{(n)}(v,x)\exp(ia_n(v)x)\sum_{k=0}^{\infty}e^{2\pi i k a_n(v)}dx
\nonumber \\
&& =\int\limits_{-1}^1 \frac{dv}{1-e^{2\pi i  a_n(v)}+i0}
\int\limits_{0}^{2\pi} F_i^{(n)}(v,x)\exp(ia_n(v)x)dx.
 \label{a9}
\end{eqnarray}
We will use in Eq.(\ref{a9}) the known equality
\begin{equation}
\frac{1}{1-e^{2\pi i a_n(v)}+i0}=\frac{\cal P}{1-e^{2\pi i
a_n(v)}}-i\pi\delta(1-e^{2\pi i a_n(v)}).\label{a10}
\end{equation}
By virtue of the made above observation
\begin{equation}
-i\pi\delta(1-e^{2\pi i a_n(v)})=-i\pi\sum_m \delta(1-e^{2\pi i
(a_n(v)-m)})\rightarrow \frac{1}{2}\sum_{m \geq
n}\delta(a_n(v)-m).\label{a11}
\end{equation}
Substituting Eq.(\ref{a10}) and Eq.(\ref{a11}) into Eq.(\ref{a9})
and taking into account that
\begin{equation}
F_i^{(n)}(v,x+\pi)=(-1)^n F_i^{(n)}(v,x),\label{a12}
\end{equation}
we have
\begin{eqnarray}
&& T_i^{(n)}=(-1)^n \frac{i}{2}{\cal
P}\int\limits_{-1}^1\frac{dv}{\sin (\pi a_n(v))}
\int\limits_{-\pi}^{\pi} F_i^{(n)}(v,x)\exp(ia_n(v)x)dx
\nonumber \\
&& +\sum_{m \geq n}\sum_{v_{1,2}}\frac{1+(-1)^{n+m}}{2|a_n'(v_)|}
\int\limits_{-\pi}^{\pi} F_i^{(n)}(v_{1,2},x)\exp(imx)dx,
 \label{a13}
\end{eqnarray}
where
\begin{eqnarray}
&&
v_{1,2}=\frac{n\mu}{2r}\pm\sqrt{\frac{n^2\mu^2}{4r^2}+1-\frac{1+m\mu}{r}},
\nonumber \\
&& |a_n'(v_)|=\frac{2}{\mu}\sqrt{r g(n,m,r)},\quad
g(n,m,r)=r-1-m\mu+\frac{n^2\mu^2}{4r}.
 \label{a14}
\end{eqnarray}
It is seen from Eq.(\ref{a13}) that $n+m$ is even. Using
Eq.(\ref{a12}) one can represent the imaginary part of $T_i^{(n,m)}$
in the form
\begin{equation}
{\rm Im}T_i^{(n,m)}= -i\sum_{v_{1,2}}\frac{\mu}{\sqrt{r
g(n,m,r)}}\int\limits_{-\pi/2}^{\pi/2}
F_i^{(n)}(v_{1,2},x)\exp(imx)dx\label{a15}
\end{equation}
Along with integers $m$ and $n$ we will use also $l=(m+n)/2$ and
$k=(m-n)/2$.

We turn now to calculation of integrals in Eq.(\ref{a15}). Let us
consider the integral
\begin{equation}
C_{n+1}^{m}(z)=(-i)^{n+1}\int\limits_{-\pi/2}^{\pi/2}
J_{n+1}\left(\frac{z}{\sin x}\right)e^{i(mx+z\cot x)}\frac{dx}{\sin
x}\label{a16}
\end{equation}
Changing the variable $t=\cot x$ we get
\begin{equation}
C_{n+1}^{m}(z)=(-i)^{n+1}\int\limits_{-\infty}^{\infty}
J_{n+1}\left(z\sqrt{t^2+1}\right)\left(\frac{t+i}{t-i}\right)^{m/2}e^{izt}\frac{dt}
{\sqrt{t^2+1}}. \label{a17}
\end{equation}
One can close the integration contour in Eq.(\ref{a17}) by an
infinite half-circle in the upper half-plane and than contract the
contour to the point $t=i$. So, we have to calculate the integral in
Eq.(\ref{a17}) over the contour $(i+)$. Changing successive the
variables $t\rightarrow it$, $y=(t-1)/(t+1)$ we get
\begin{equation}
C_{n+1}^{m}(z)=e^{-z}\oint\limits_{(0+)}
I_{n+1}\left(\frac{2z\sqrt{y}}{1-y}\right)y^{-m/2}e^{-2zy/(1-y)}\frac{dy}
{\sqrt{y}(1-y)}. \label{a18}
\end{equation}
Using Eq.10.12.(20) \cite{BE}
\begin{equation}
\frac{1}{1-y} e^{-2zy/(1-y)}
(z\sqrt{y})^{-n}I_{n}\left(\frac{2z\sqrt{y}}{1-y}\right)=\sum_{k=0}^{\infty}
\frac{k!}{(k+n)!}(L_{k}^{n}(z))^2 y^k, \label{a19}
\end{equation}
where $I_{n}(x)$ is the modified Bessel function,  $L_{k}^{n}(z)$ is
the is the generalized Laguerre polynomial, we get
\begin{equation}
C_{n+1}^{m}(z)=2\pi i e^{-z} z^{n+1} \frac{(k-1)!}{l!}
(L_{k-1}^{n+1}(z))^2 \vartheta(k-1), \quad n=l-k. \label{a20}
\end{equation}
In the same manner one can calculate the integral
\begin{eqnarray}
\hspace{-10mm}&& S_n^m(z)=2(-i)^n\int\limits_{-\pi/2}^{\pi/2} \cot x
J_{n}\left(\frac{z}{\sin x}\right)e^{i(mx+z\cot x)}dx
\nonumber \\
\hspace{-10mm}&& =2\pi i e^{-z} z^{n} \frac{k!}{l!}F(l,k,z),\quad
F(l,k,z)=(L_{k}^{n}(z))^2+\frac{l}{k}(L_{k-1}^{n}(z))^2\vartheta(k-1).
 \label{a21}
\end{eqnarray}
Somewhat different structure has the integral
\begin{eqnarray}
\hspace{-18mm}&& D_n^m(z)=(-i)^n\int\limits_{-\pi/2}^{\pi/2}
J_{n}\left(\frac{z}{\sin x}\right)e^{i(mx+z\cot x)}\frac{dx}{\sin^2
x}
\nonumber \\
\hspace{-18mm}&& =2 i e^{-z} \oint\limits_{(0+)}
I_{n+1}\left(\frac{2z\sqrt{y}}{1-y}\right)y^{-m/2}e^{-2zy/(1-y)}\frac{dy}
{(1-y)^2}
\nonumber \\
\hspace{-18mm}&&= 2 i e^{-z} z^n
\oint\limits_{(0+)}\sum_{s=0}^{\infty}\sum_{j=0}^{\infty}\frac{j!}{(j+n)!}
(L_{j}^{n}(z))^2 y^{j+s-k} = -4 \pi e^{-z} z^n
\sum_{j=0}^{k-1}\frac{j!}{(j+n)!} (L_{j}^{n}(z))^2.
 \label{a22}
\end{eqnarray}
Using Eq.10.12.(9) \cite{BE}:
\begin{equation}
\sum_{j=0}^{k-1}\frac{j!}{(j+n)!}
L_{j}^{n}(x)L_{j}^{n}(y)=\frac{k!}{(k+n-1)!}\frac{1}{x-y}
\left[L_{k-1}^{n}(x)L_{k}^{n}(y)-L_{k}^{n}(x)L_{k-1}^{n}(y)\right],
\label{a23}
\end{equation}
one obtains
\begin{eqnarray}
\hspace{-10mm}&& D_n^m(z)=-4 \pi e^{-z} z^n
\frac{k!}{(l-1)!}\left[L_{k}^{n}(z)\frac{d}{dz}L_{k-1}^{n}(z)
-L_{k-1}^{n}(z)\frac{d}{dz}L_{k}^{n}(z)\right]
\nonumber \\
\hspace{-10mm}&&=4 \pi e^{-z} z^n
\frac{k!}{l!}l\left[L_{k-1}^{n+1}(z)L_{k}^{n-1}(z)
-L_{k}^{n}(z)L_{k-1}^{n}(z)\right].
 \label{a24}
\end{eqnarray}
In derivation of the last expression Eqs.10.12.(15),(16),(24)
\cite{BE} was used. In conclusion of this Appendix let us note, that
$T_i^{(n,m)}$ contains the combination
\begin{eqnarray}
\hspace{-10mm}&& C_{n+1}^m(z)+C_{n-1}^m(z)=2 \pi i e^{-z} z^{n-1}
\frac{k!}{l!}\left[\frac{z^2}{k}(L_{k-1}^{n+1}(z))^2
+l(L_{k}^{n-1}(z))^2\right]
\nonumber \\
\hspace{-10mm}&&=2 \pi i e^{-z} z^{n-1}
\frac{k!}{l!}\left[\frac{1}{k}(lL_{k-1}^{n}(z)-kL_{k}^{n}(z))^2
+l(L_{k}^{n}(z)-L_{k-1}^{n}(z))^2\right]
\nonumber \\
\hspace{-10mm}&&=2 \pi i e^{-z} z^{n-1} \frac{k!}{l!}\left[m
F(l,k,z) - 4lL_{k}^{n}(z)L_{k-1}^{n}(z)\right].
 \label{a25}
\end{eqnarray}
It should be noted that the above derivation is the most direct path
from the process probability in the form of the integral
Eqs.(\ref{d1}),(\ref{d2}) to the process probability in the form
Eq.(\ref{d7}).

\setcounter{equation}{0}

\section{Appendix}

{\large{\bf Quasiclassical approximation at low photon energy
\newline ($\omega \sim m,~k^2=0,~r=q$)}}

\vskip2mm

In the field which is weak comparing with the critical field
$B/B_0=\mu \ll 1$ and at relatively low photon energy ($r \sim 1$)
the created particles occupy mainly states with high quantum numbers
if the condition $r-1 \gg \mu$ is fulfilled. In this case the
quasiclassical approach is valid, but created particles are no more
ultrarelativistic, so the standard quasiclassical formulas
\cite{BK,BKF,BKS1} are nonapplicable. We will develop here another
approach, using the method of stationary phase at calculation of the
imaginary part of the integral over $x$ in Eq.(\ref{d2}). Granting
that the large parameter $1/\mu$ is the common factor in the phase
$\psi(x)$ and doesn't contained in the equation $\psi'(x)=0$ which
defines the stationary phase point $x_0(r \sim 1)\sim 1$. In this
case the small values of variable $v$ contribute to the integral
over $v$, so that one can extend the integration limits to the
infinity. So we get
\begin{equation}
{\rm Im}\Omega_i \simeq i\frac{\alpha
m^2}{2\pi}\int\limits_{-\infty}^{\infty}dv\int\limits_{-\infty}^{\infty}
f_i(v,x)\exp\left\{-\frac{i}{\mu}
\left[\varphi(x)+v^2\chi(x)\right]\right\}dx,\label{b1}
\end{equation}
where
\begin{equation}
\varphi(x)=2r\tan\left(\frac{x}{2}\right)+(1-r)x,\quad
\chi(x)=rx\left(1-\frac{x}{\sin x}\right).\label{b2}
\end{equation}
From the equation $\varphi'(x)=0$ we find
\begin{equation}
\tan\left(\frac{x_0}{2}\right)=-\frac{i}{\sqrt{r}}, \quad
x_0(r)=-il(r), \quad
l(r)=\ln\frac{\sqrt{r}+1}{\sqrt{r}-1}.\label{b3}
\end{equation}
Substituting these results in the expressions which defines the
integral in Eq.(\ref{b1}) we have
\begin{eqnarray}
\hspace{-10mm}&& i\varphi(x_0)=\beta(r)=2\sqrt{r}-(r-1)l(r),\quad
i\varphi''(x_0)=\frac{r-1}{\sqrt{r}}, \quad
i\chi(x_0)=\frac{\sqrt{r}}{2}l(r)\beta(r)
\nonumber \\
\hspace{-10mm}&& if_2(v=0, x_0(r))=\frac{r-1}{2r^{3/2}}, \quad
-if_3(v=0, x_0(r))=\frac{1}{\sqrt{r}}.
 \label{b4}
\end{eqnarray}
Performing the standard procedure of the method of stationary phase
one obtains for probability of pair creation by a polarized photon
not far from threshold
\begin{equation}
W_3^{(th)}=\frac{\alpha
m^2\mu}{\omega}\sqrt{\frac{r}{(r-1)l(r)\beta(r)}}
\exp\left(-\frac{\beta(r)}{\mu}\right),\quad
W_2^{(th)}=\frac{r-1}{2r}W_3^{(th)}. \label{b5}
\end{equation}
In spite of made above assumption $r \sim 1$ Eq.(\ref{b5}) is valid
also at $r \gg 1$ if the condition $\beta(r) \gg \mu$ is fulfilled.
This can be traced in the derivation of Eq.(\ref{b5}). The first two
term of the decomposition of the function $\beta(r)$ over power of
$1/r$ are
\begin{equation}
\frac{\beta (r)}{\mu}\simeq \frac{4}{3\mu\sqrt{r}}+\frac{4}{15\mu
r^{3/2}}\label{b6}
\end{equation}
It follows from this formula that that applicability of
Eq.(\ref{b5}) is limited by the condition $r \ll \mu^{-2}$.  If the
second term much smaller than unity the exponent with it can be
expanded. As a result we have from Eq.(\ref{b5}) at $\mu^{-2/3}\ll r
\ll \mu^{-2}$
\begin{equation}
W_3=\frac{\alpha m^2\mu}{2\omega}\sqrt{\frac{3r}{2}}
\exp\left(-\frac{4}{3\mu\sqrt{r}}\right)\left(1- \frac{4}{15\mu
r^{3/2}}\right), \quad W_2=\frac{1}{2}W_3\label{b7}
\end{equation}
The main term in above expression coincides with the probability of
pair creation by a photon in standard quasiclassical theory at
$\kappa=2\mu\sqrt{r} \ll 1$. The correction in Eq.(\ref{b7})
determines the lower boundary over the photon energy of standard
approach applicability ($\kappa^3 \gg \mu^2$). So exists the
overlapping region where both the formulated here and the standard
approach for high energy are valid.

\setcounter{equation}{0}

\section{Appendix}

{\large{\bf Corrections to the standard quasiclassical approximation
\newline ($k^2=0,~r=q$)}}
\vskip2mm

The standard quasiclassical approximation valid for
ultrarelativistic created particles ($r \gg 1$) can be derived from
Eqs.(\ref{d2}), (\ref{d3}) by expanding the functions $f_2(v,x),
f_3(v,x), \psi(v,x)$ over $x$ powers. Taking into account the higher
powers of $x$ one gets
\begin{eqnarray}
\hspace{-10mm}&&
f_2(v,x)=\frac{1-v^2}{12}\left[(3+v^2)x+\frac{1}{15}(15-6v^2-v^4)x^3\right]
\nonumber \\
\hspace{-10mm}&&
f_3(v,x)=-\frac{1-v^2}{6}\left[(3-v^2)x+\frac{1}{60}(15-2v^2+3v^4)x^3\right]
\nonumber \\
\hspace{-10mm}&&
\psi(v,x)=-\frac{r(1-v^2)^2}{12\mu}\left(x^3+\frac{3-v^2}{30}x^5\right)
-\frac{x}{\mu}.
 \label{c1}
\end{eqnarray}
Here the first terms in the brackets give the known probability of
the process in the standard quasiclassical approximation, while the
second terms are the corrections. Expanding the term with $x^5$ in
$\exp(i\psi(v,x))$ and making substitution $x=\mu t$ one finds
\begin{eqnarray}
\hspace{-15mm}&& {\rm Im}\Omega_i = i\frac{\alpha
m^2\mu}{2\pi}\int\limits_{-1}^{1}dv\int\limits_{-\infty}^{\infty}
g_i(v,t)\exp\left[-i\left(t+\xi\frac{t^3}{3}\right)\right]dx,
\nonumber \\
\hspace{-15mm}&& g_2(v,t)=\frac{1-v^2}{12}\mu
t\left[(3+v^2)-i\frac{9-v^4}{90}\xi\mu^2t^5+
\frac{\mu^2t^2}{15}(15-6v^2-v^4)\right],
\nonumber \\
\hspace{-15mm}&& g_3(v,t)=-\frac{1-v^2}{6}\mu
t\left[(3-v^2)-i\frac{(3-v^2)^2}{90}\xi\mu^2t^5+
\frac{\mu^2t^2}{60}(15-2v^2+3v^4)\right],
 \label{c2}
\end{eqnarray}
where
\begin{equation}
\xi=\frac{[(1-v^2)\kappa]^2}{16},\quad
\kappa=2\sqrt{r}\mu=\frac{\omega}{m}\mu. \label{c3}
\end{equation}
We will use the known integrals
\begin{eqnarray}
\hspace{-5mm}&&
\int\limits_{-\infty}^{\infty}\cos\left(t+\xi\frac{t^3}{3}\right)
=\sqrt{3}z K_{1/3}(z),\quad
z=\frac{2}{3\sqrt{\xi}}=\frac{8}{3(1-v^2)\kappa},
\nonumber \\
\hspace{-5mm}&&
\int\limits_{-\infty}^{\infty}t\sin\left(t+\xi\frac{t^3}{3}\right)
=\frac{3\sqrt{3}}{2}z^2 K_{2/3}(z).
 \label{c4}
\end{eqnarray}
Differentiating the first integral over $\xi$ one gets
\begin{eqnarray}
\hspace{-1mm}&&
\int\limits_{-\infty}^{\infty}t^3\sin\left(t+\xi\frac{t^3}{3}\right)
=\frac{27\sqrt{3}}{8}z^3\frac{d}{dz}\left(z K_{1/3}(z)\right)
=-\frac{27\sqrt{3}}{8}z^3\left(zK_{2/3}(z)-\frac{2}{3}K_{1/3}(z)\right),
\nonumber \\
\hspace{-1mm}&&
\xi\int\limits_{-\infty}^{\infty}t^6\cos\left(t+\xi\frac{t^3}{3}\right)
=-\frac{81\sqrt{3}}{16z^2}\left(z^3\frac{d}{dz}\right)^2\left(z
K_{1/3}(z)\right)
\nonumber \\
\hspace{-1mm}&& =\frac{81\sqrt{3}}{16}z^3\left[4zK_{2/3}(z)-
\left(\frac{16}{9}+z^2\right)K_{1/3}(z)\right].
 \label{c5}
\end{eqnarray}
Substituting the integrals Eqs.(\ref{c3})-(\ref{c4}) in
Eq.(\ref{c2}) we obtain the probabilities of pair creation in
standard quasiclassical approximation
\begin{eqnarray}
\hspace{-5mm}&& W_2=-{\rm Im}\frac{\kappa_2}{\omega} =\frac{\alpha
m^2}{3\sqrt{3}\pi\omega}
\int\limits_{-1}^{1}\frac{3+v^2}{1-v^2}K_{2/3}(z)dv,
\nonumber \\
\hspace{-5mm}&& W_3=-{\rm Im}\frac{\kappa_3}{\omega} =\frac{2\alpha
m^2}{3\sqrt{3}\pi\omega}
\int\limits_{-1}^{1}\frac{3-v^2}{1-v^2}K_{2/3}(z)dv,
 \label{c6}
\end{eqnarray}
and the corresponding corrections to this approximation
\begin{equation}
W_i^{(1)}=\frac{\alpha m^2\mu^2}{30\sqrt{3}\pi\omega\kappa}
\int\limits_{0}^{1}G_i(v,z)\frac{dv}{1-v^2}, \label{c7}
\end{equation}
where
\begin{eqnarray}
&& G_2(v,z)=(36+4v^2-18z^2)K_{1/3}(z)+(3v^2-57)zK_{2/3}(z),
\nonumber \\
&& G_3(v,z)=-(34+2v^2+36z^2)K_{1/3}(z)+(78-6v^2)zK_{2/3}(z).
 \label{c8}
\end{eqnarray}
In derivation of Eqs.(\ref{c7})-(\ref{c8}) the integration by parts
was used as well as the intermediate equalities in Eq.(\ref{c5}).
The obtained representation of probabilities is convenient for
calculation asymptotic at $\kappa \gg 1$ which are
\begin{eqnarray}
&& W_i^{(1)}=\frac{\alpha m^2\mu^2}{30\sqrt{3}\pi\omega\kappa}w_i,
\quad w_2=12 A \kappa^{1/3}-90\pi,
\nonumber \\
&&  w_3=-11 A \kappa^{1/3}+84\pi,\quad
A=3^{1/3}\frac{2}{5}\frac{\Gamma^3(1/3)}{\Gamma(2/3)} = 8.191.. .
 \label{c9}
\end{eqnarray}
At $\kappa \ll 1$ the main contribution is given by the terms
$\propto z^2$ in $G_i$~ Eq.(\ref{c8}). One has
\begin{equation}
W_2^{(1)}=\frac{\alpha
m^2\mu^2}{\omega\kappa^2}\frac{2\sqrt{2}}{5\sqrt{3}}
\exp\left(-\frac{8}{3\kappa}\right), \quad W_3^{(1)}=2W_2^{(1)}.
 \label{c10}
\end{equation}
The relative magnitude of the corrections at $\kappa \gg 1$
\begin{equation}
\frac{ W_2^{(1)}}{W_2} \simeq
\frac{2\mu^2}{\kappa^{4/3}}\left(1-\frac{2.9}{\kappa^{1/3}}\right),\quad
\frac{ W_3^{(1)}}{W_3} \simeq
\frac{11\mu^2}{9\kappa^{4/3}}\left(\frac{2.9}{\kappa^{1/3}}-1\right),
 \label{c11}
\end{equation}
At $\kappa \ll 1$ one has
\begin{equation}
\frac{W_i^{(1)}}{ W_i}  \simeq -\frac{32\mu^2}{15\kappa^3},
 \label{c12}
\end{equation}
what agrees with Eq.(\ref{b7}).

\setcounter{equation}{0}

\section{Appendix}

{\large{\bf On the derivation of created particle spectrum in SQA}}
\vskip2mm

 Substituting in Eq.(\ref{s3}) $x=(1+v)/2$ one has
\begin{equation}
\frac{dW^{SQA}}{dv}=\frac{\alpha m^2}{2\sqrt{3}\pi \omega}
\left[2\frac{1+v^2}{1-v^2}K_{2/3}(z)+
\int_{z}^{\infty}K_{1/3}(y)dy\right],
 \label{ad1}
\end{equation}
where $z$ is defined in Eq.(\ref{d13}). If one omits the integration
over $v$ in Eq.(\ref{d18}), the result of this manipulation
disagrees with Eq.(\ref{ad1}). So the described manipulation is
erroneous. Only if one performs integration by parts of the second
term in Eq.(\ref{ad1}) and use the recursion relation
\begin{equation}
zK_{1/3}(z)=-z\frac{dK_{2/3}(z)}{dz}-\frac{2}{3}K_{2/3}(z),
 \label{ad2}
\end{equation}
one obtains Eq.(\ref{d18}) for integral probability.

\clearpage

\begin{figure}[p]
\begin{picture}(10,120)
\put(70,160){\includegraphics[height=11cm]{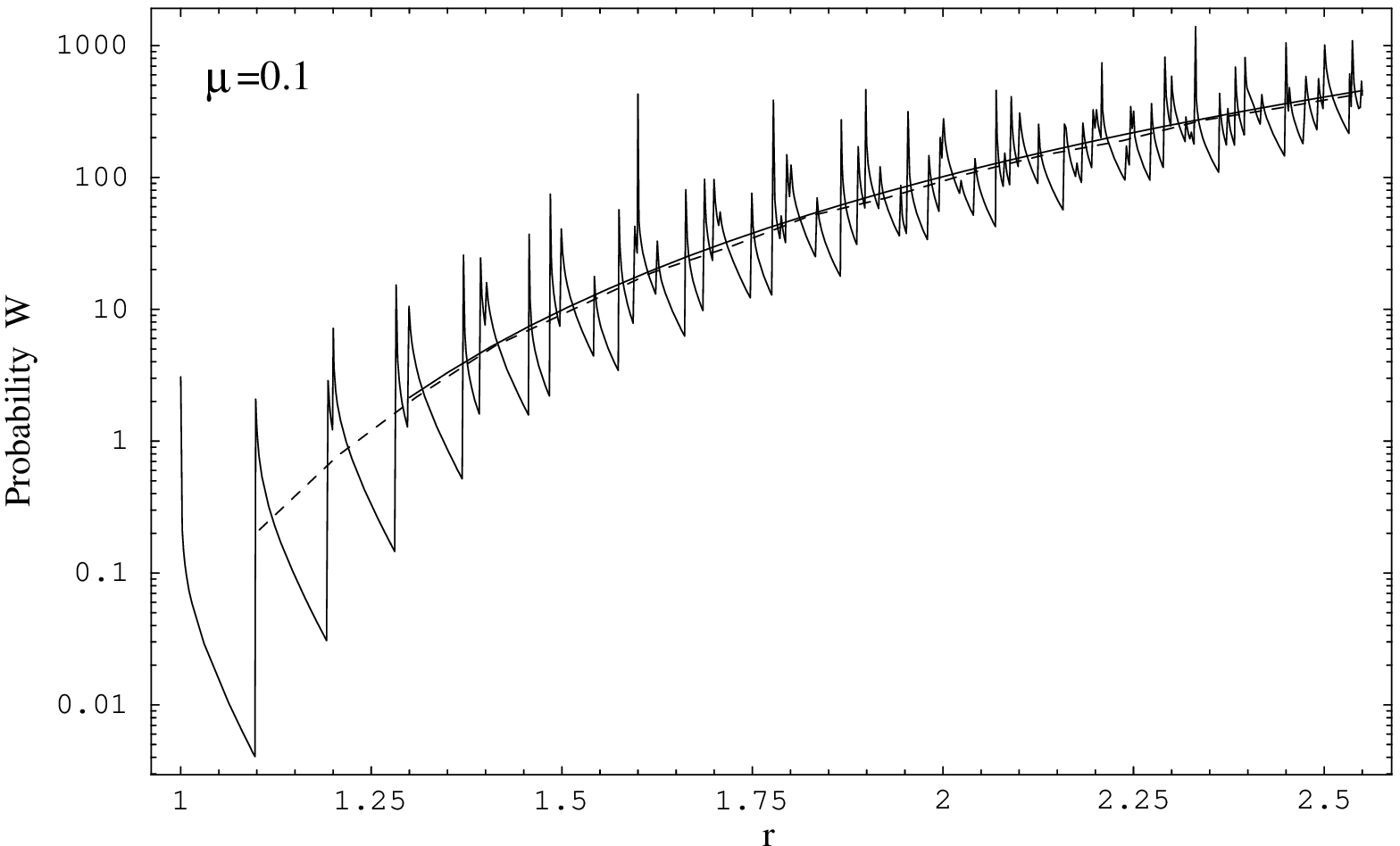}}
\end{picture}
\caption{The pair creation probability for unpolarized photon at
$\mu=0.1$ as a function of $r$. The probability in saw-tooth form is
calculated according to Eq.(\ref{d10}). The dashed line is the
result of averaging over the interval $(-\mu/2+r \div \mu/2+r)$. The
solid line is calculated according to Eq.(\ref{d15}).}
\end{figure}

\clearpage

\begin{figure}[p]
\begin{picture}(0,200)
\put(-40,8){\includegraphics[height=10cm]{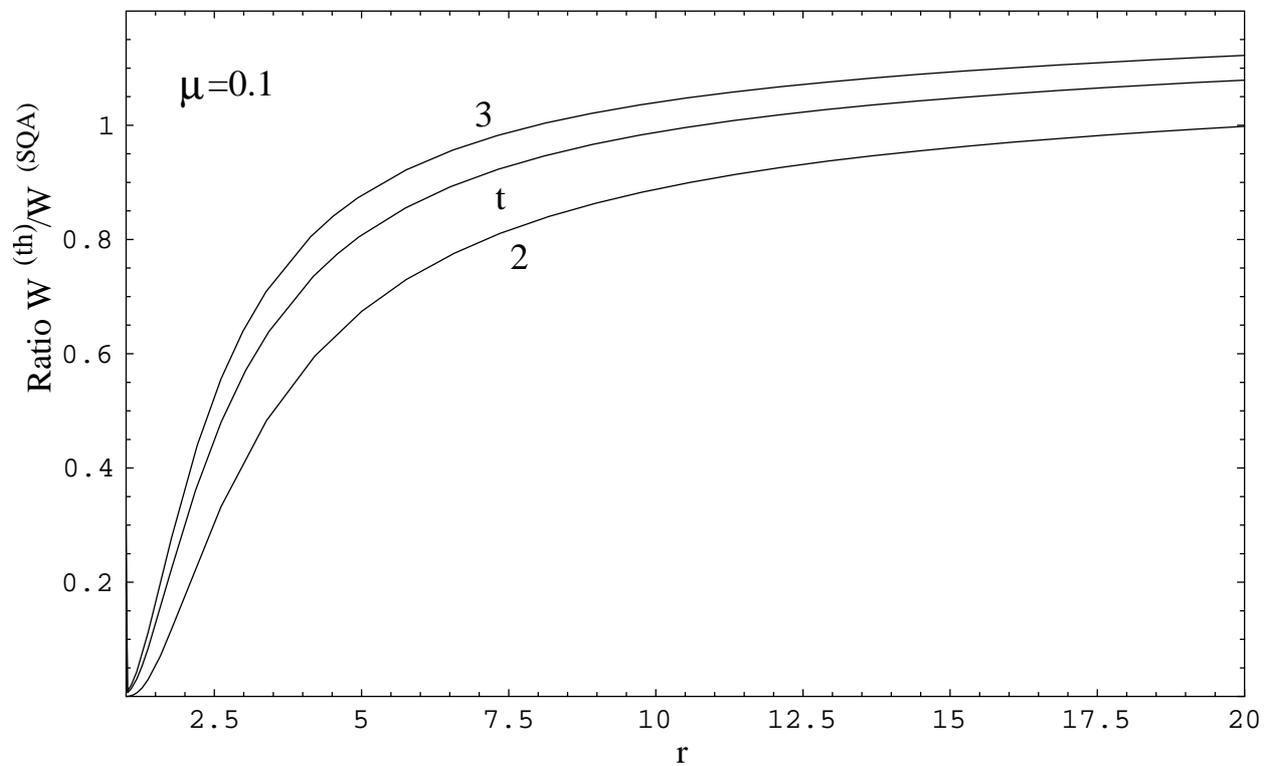}}
\end{picture}
\caption{The ratio of probabilities $W^{(th)}/W^{(SQA)}$. The curves
2 and 3 are for polarized photons, the curve t is for unpolarized
photon (see Eq.(\ref{d15}) and Eq.(\ref{d18})). }
\end{figure}

\clearpage

\begin{figure}[p]
\begin{picture}(0,200)
\put(-80,8){\includegraphics[height=10cm]{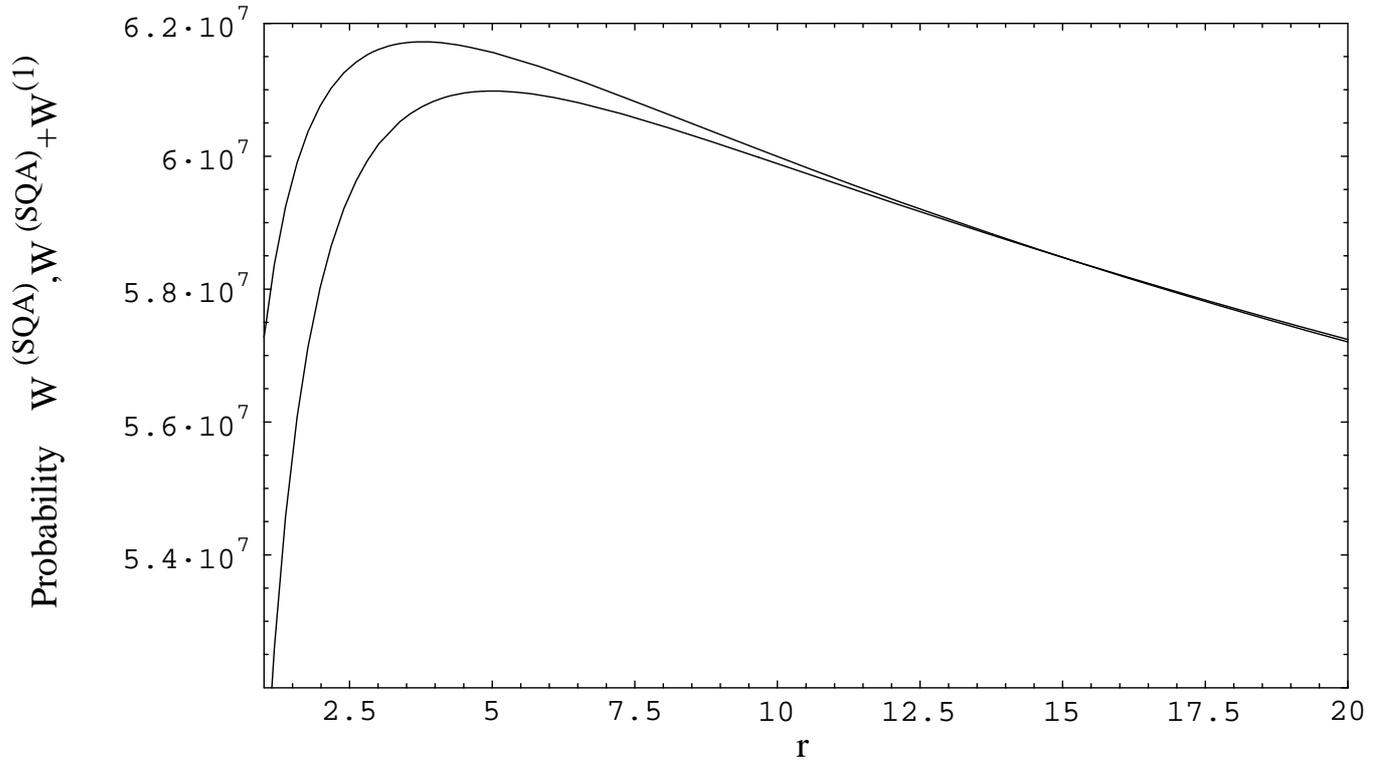}}
\end{picture}

\caption{The pair creation by a photon probability at $\mu=3$ in
quasiclassical approximation. The upper curve is the probability in
SQA Eq.(\ref{d18}). The lower curve is the sum $W^{(SQA)}+W^{(1)}$
Eq.(\ref{d12})}
\end{figure}

\clearpage

\begin{figure}[p]
\begin{picture}(0,200)
\put(-65,8){\includegraphics[height=10cm]{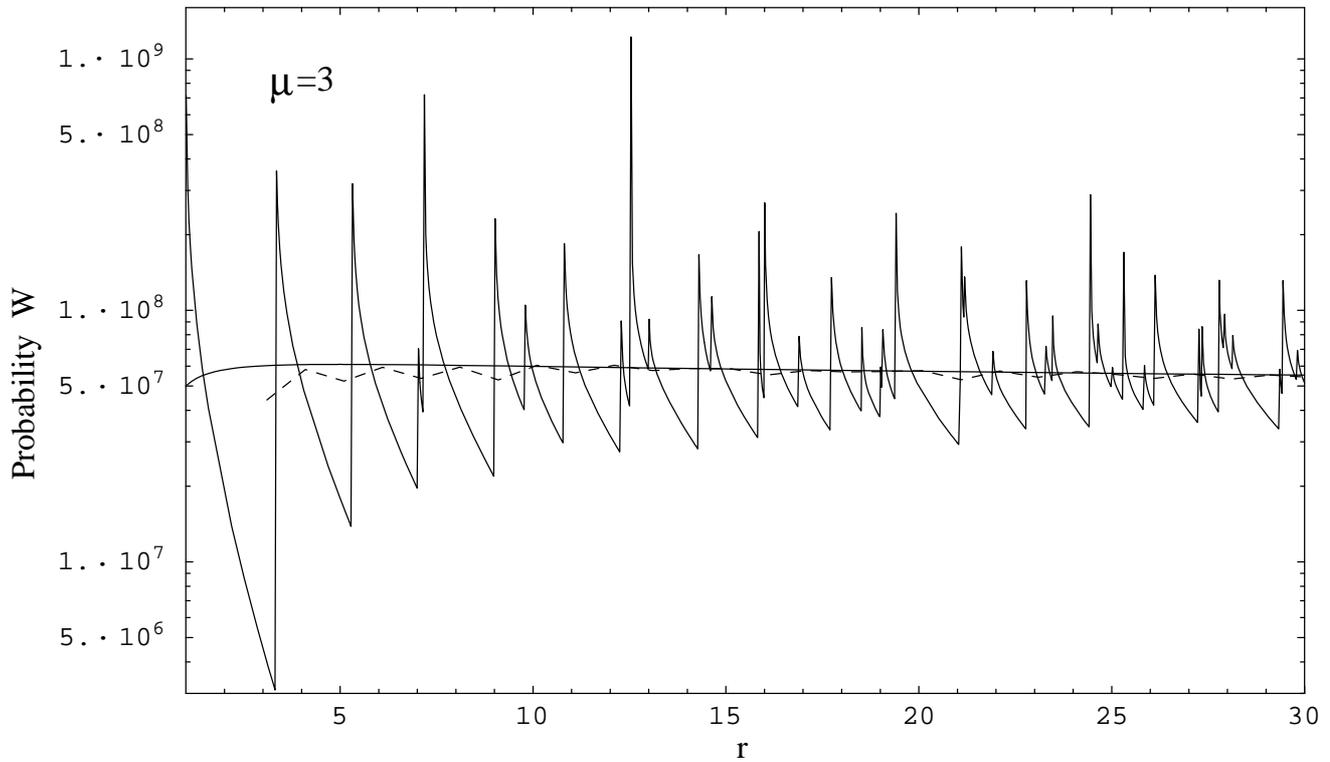}}
\end{picture}

\caption{The probability of pair creation at $\mu=3$. The
probability in saw-tooth form is calculated according to
Eq.(\ref{d10}).  The result of averaging over the interval
$(-\mu/2+r \div \mu/2+r)$ is given by the dashed line. The solid
curve, which is quasiclassical approximation, calculated according
to Eqs.(\ref{d18}) and (\ref{d12}).}
\end{figure}

\clearpage

\begin{figure}[p]
\begin{picture}(0,200)
\put(-65,8){\includegraphics[height=10cm]{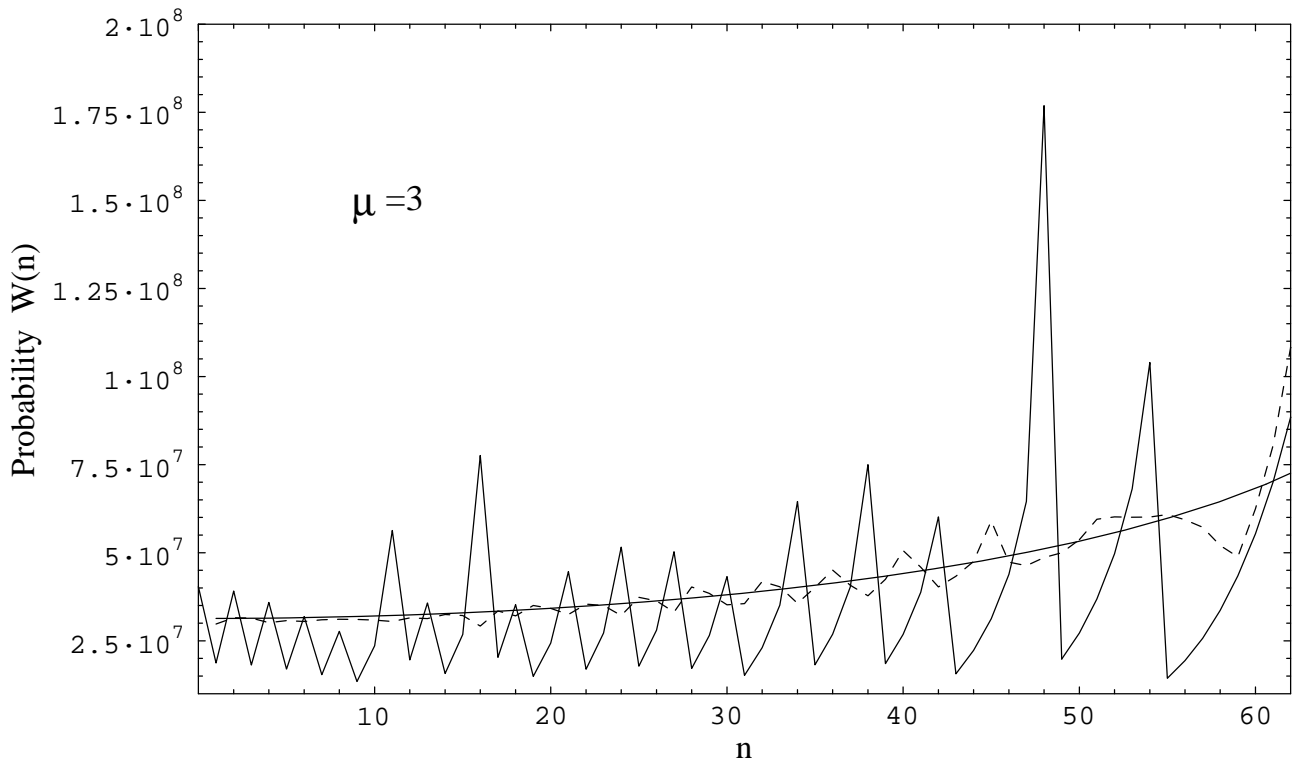}}
\end{picture}

\caption{The spectrum of created particles for
$\mu=3,~r=110~(n_m=66)$ vs $n$ (see Eq.(\ref{s2})). The spectrum in
saw-tooth form is calculated according to Eq.(\ref{d10}). The dashed
curve is the spectrum averaged over the interval $(-2\mu+r \div
2\mu+r)$. The solid curve calculated according to Eq.(\ref{s3}). }
\end{figure}

\clearpage

\begin{figure}[p]
\begin{picture}(0,200)
\put(-60,8){\includegraphics[height=10cm]{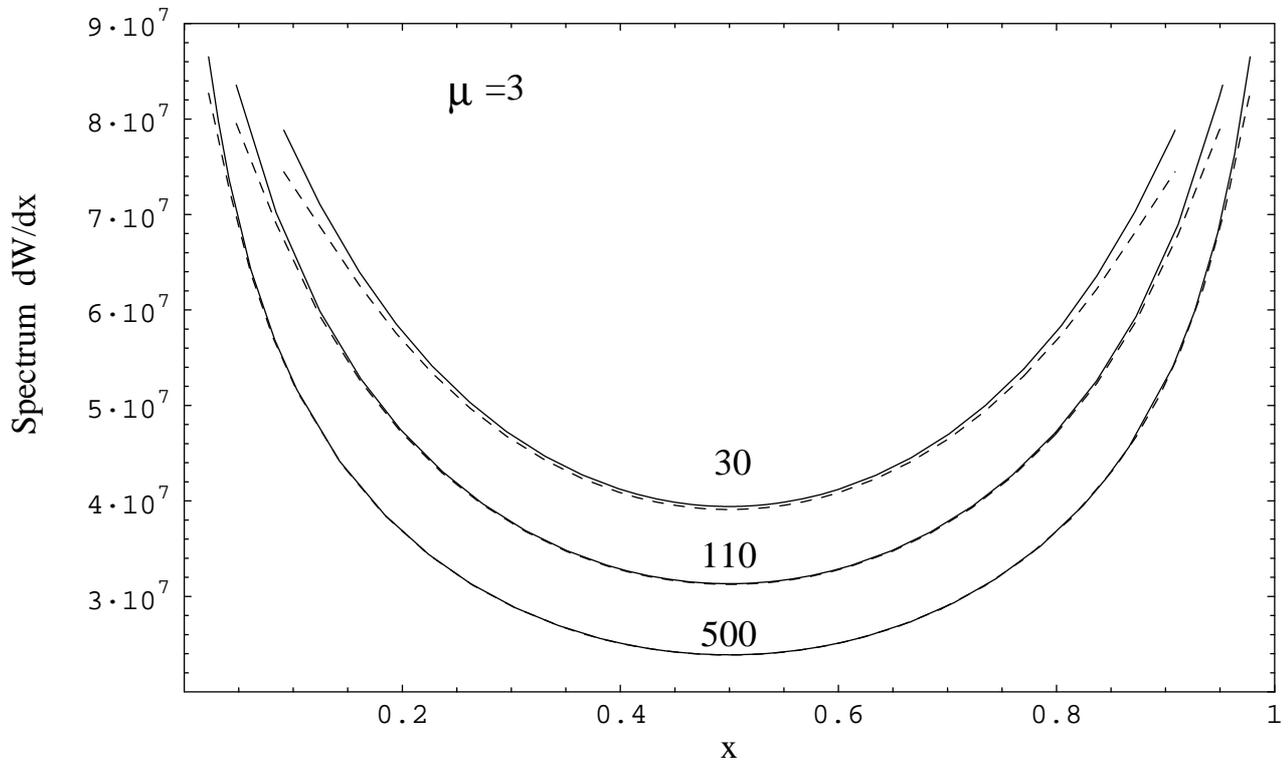}}
\end{picture}

\caption{The energy distribution of the created electron(positron)
at $\mu=3$ for different r: $r=30,~n_m=16$ (the upper curve),
$r=110,~n_m=66$ (the middle curve), $r=500,~n_m=318$ (the lower
curve), calculated according to Eq.(\ref{s3}) (the solid curve) and
according to Eq.(\ref{s6}) (the dashed curve).}
\end{figure}

\end{document}